\documentclass[twocolumn]{aa}

\usepackage{color}
\usepackage{epsfig}
\usepackage{soul}
\usepackage{float}
\usepackage{enumitem}
\usepackage{appendix}
\usepackage{wrapfig}
\usepackage{graphicx}
\usepackage{multicol}
\usepackage{hyperref}
\usepackage{chemfig}
\usepackage{longtable}
\usepackage{supertabular,booktabs}
\usepackage{pdflscape}
\usepackage{longtable}
\usepackage{textcomp}
\usepackage{amssymb}
\definecolor{Taurus}{HTML}{006D77}
\definecolor{B1}{HTML}{5C3D73}
\definecolor{IC348}{HTML}{922D50}
\definecolor{B5}{HTML}{DAA520}

\usepackage[version=3]{mhchem}
\hypersetup{
    colorlinks=true,
    linkcolor=blue,
    filecolor=magenta,
    citecolor=blue,
    urlcolor=blue,
    pdftitle={Overleaf Example},
    pdfpagemode=FullScreen,
    }
\raggedcolumns
\usepackage{natbib}

\def\aap{A\&A}
\def\apjs{ApJS}

\usepackage{blindtext}
\usepackage{lipsum}
\usepackage{afterpage}

\newsavebox{\splitcaptionbox}
\newlength{\splitcaptionheight}
\makeatletter
\long\def\@makesplitcaption#1#2{
  \if@stage@final
    \vskip0.7\abovecaptionskip
    \addtolength{\splitcaptionheight}{-0.7\abovecaptionskip}
  \else
    \vskip\abovecaptionskip\goodbreak
    \addtolength{\splitcaptionheight}{-\abovecaptionskip}
  \fi
  \setbox\splitcaptionbox=\vbox{\interlinepenalty 0
    \reset@font\small{\bfseries#1.} #2\par}
  \vsplit\splitcaptionbox to \splitcaptionheight
  \global\setbox\splitcaptionbox=\box\splitcaptionbox
  \if@cop@home\ifonline\ifnum\csname c@\@captype\endcsname=1 
    \immediate\write\@auxout{\string\gdef\string\@num\@captype{}}
    \hypertarget{\@captype}{}
  \fi\fi\fi}

\newcommand{\splitcaption}[3][\empty]
 {\setlength{\splitcaptionheight}{#3}
  \let\cop@makecaption=\@makecaption
  \let\@makecaption=\@makesplitcaption
  \ifx\empty#1\relax
    \caption{#2}
  \else
    \caption[#1]{#2}
  \fi
  \let\@makecaption=\cop@makecaption}
\makeatother

\newcommand{\mergecaption}{\ifdim\ht\splitcaptionbox>0pt
  \begin{figure}[tp]
  \box\splitcaptionbox
  \end{figure}
\fi}

\begin{document}

\title{Gas phase Elemental abundances in Molecular cloudS (GEMS)}
\subtitle{XII. First look into the para-to-ortho ratio of c-C$_{3}$D$_{2}$ towards prestellar cores}

\author{J.~Ferrer Asensio, \inst{1} A.~Fuente,\inst{2} E.~Roueff,\inst{3} K.~Furuya,\inst{1} B.~Tercero,\inst{4,5} N.~Sakai\inst{1}} 
\institute{\tiny{\inst{1} RIKEN Cluster for Pioneering Research, Wako-shi, Saitama, 351-0106, Japan\\ \inst{2} Centro de Astrobiolog\'ia (CAB), CSIC-INTA, Carretera de Ajalvir, km 4, E-28805 Torrej\'on de Ardoz, Spain\\ 
\inst{3} LUX, Observatoire de Paris, PSL Research University, CNRS, Sorbonne Universit\'es, 92190 Meudon, France\\ \inst{4} Observatorio Astron\'omico Nacional (IGN), C/ Alfonso XII 3, 28014 Madrid, Spain}\\ \inst{5} Observatorio de Yebes (IGN), Cerro de la Palera s/n, 19141 Yebes, Guadalajara, Spain}

\titlerunning{GEMS: XII. First look into the para-to-ortho ratio of c-C$_{3}$D$_{2}$ towards prestellar cores}
\authorrunning{J. Ferrer Asensio et al.}

\date{Received  ; accepted }

\abstract
{Symmetric molecules containing several identical nuclei give rise to distinct isomers characterized by the total nuclear spin and producing separate ortho and para species that can not be exchanged through radiative or inelastic transitions and whose relative populations can preserve information about formation conditions. In the interstellar medium, cyclopropenylidene (c‑C$_{3}$H$_{2}$) and its doubly-deuterated isotopologue provide a prime laboratory for studying spin isomer ratios, which have been measured for the main species but are yet unavailable for c‑C$_{3}$D$_{2}$.}
{We present new detections of both ortho and para transitions of c-C$_{3}$D$_{2}$ towards three sources in the Taurus molecular cloud from the  GEMS Large program, and revisit previous detections towards four other cores from the literature in order to investigate for the first time the para-to-ortho ratio  for doubly deuterated c-C$_{3}$D$_{2}$.}
{The ortho and para c-C$_{3}$D$_{2}$ column densities towards all the sources are computed using the non-local thermodynamic equilibrium (non-LTE) software RADEX thanks to the recent availability of dedicated collision rate coefficients.}
{The total derived column densities of c-C$_{3}$D$_{2}$ for the new sources present  a range of (2.5 - 5.3) $\times$ 10$^{11}$ cm$^{-2}$. The para-to-ortho ratio derived for B213-C16-1 and TMC1-C, 1.5$^{+0.7}_{-0.6}$ and 1.3$^{+1.3}_{-0.7}$, respectively, are higher than the statistical ratio of 0.5, while for TMC1-NH3 the ratio is statistical within errors, 1.0$^{+1.4}_{-0.7}$. The para-to-ortho ratio for the revisited sources from the literature, L1544, TMC-1C, TMC-2 and L1517B are 0.4$^{+0.1}_{-0.1}$, 0.5$^{+0.1}_{-0.1}$, 0.6$^{+0.2}_{-0.1}$ and 1.4$^{+0.7}_{-0.5}$, respectively. In this case, the first three cores show para-to-ortho statistical values, and the latter a non-statistical one. Thus, from all the sources studied four have statistical c-C$_{3}$D$_{2}$ para-to-ortho ratios, and three sources show higher non-statistical ratios.}
{This is a first observational determination of the para-to-ortho ratio in a doubly deuterated molecule derived through a detailed radiative transfer tool. The observed larger than statistical c-C$_{3}$D$_{2}$ para-to-ortho ratio values for some of the sources may result from p-\ce{D2H+} or D$_{3}^{+}$ driven reactions. The actual ratio is not directly related to the ortho-to-para ratios of \ce{H2} and c-\ce{C3H2}. These findings provide an important reference point for future chemical models investigating the formation pathways of c-C$_{3}$D$_{2}$. } 


\maketitle

\section{Introduction} \label{introduction3}

The presence of identical nuclei in a molecule leads to an additional symmetry involving the exchange of nuclei. Hydrogen (H) and its isotopologue deuterium (D) with nuclear spins 1/2 and 1 respectively, have specific properties. The total molecular wavefunction (including electronic and nuclear variables) is antisymmetric in the case of half integer nuclear spins (fermions) and symmetric in the case of integer nuclear spins (bosons) under the exchange of nuclei geometrical operation. The different rotational levels are then linked to the actual total nuclear spin ($I_{\rm tot}$), which distinguishes the different nuclear spin isomers. In the case of two identical symmetric nuclei, the levels with a symmetric nuclear spin function are labelled as '$\textit{ortho}$' whereas the other levels are '$\textit{para}$'.\

Cyclopropenylidene (c-C$_{3}$H$_{2}$) has such properties where ortho and para levels correspond respectively to $I_{tot}=1$ and $I_{tot}=0$, with respective statistical weights 3 and 1.\footnote{The states for which the sum $K_a$ + $K_c$  is odd are ortho, and the states for which the sum $K_a$ + $K_c$ sum is even are para, where $K_a$ and $K_c$ are the projection  of the rotational quantum number $J$ on the $a$ and $c$ axes of the molecule.}
The lowest $J_{KaKc}  = 1_{01}$ ortho level is above  the ground $J_{KaKc}  = 0_{00}$ para level by a 2.5 K energy value so that the high-temperature statistical \textit{ortho}-to-\textit{para} ratio (o/p) is  3.\

Cyclopropenylidene has been detected towards many sources at different stages of the star formation process \citep{thaddeus:85, cox:88, madden:89, lucas:00, chantzos:18, giers:22, martinezhenares:25, lis:25}. Measurements of the o/p ratio of c-C$_{3}$H$_{2}$ towards starless and prestellar cores show, for some sources, a deviation from the statistical value of 3 \citep{takakuwa:01, park:06, morisawa:06, ferrerasensio:26, hsu:26}. \cite{takakuwa:01} found a o/p ratio of 2.4$^{+0.1}_{-0.1}$ towards TMC-1C. They suggest that the low o/p ratio may arise from formation and destruction processes influenced by nuclear-spin selection rules (\citealt{quack:77}, see also \citealt{oka:04}), which can in principle lead to different effective reaction pathways and rate coefficients for ortho and para species if total nuclear spin is conserved during the reaction. \cite{morisawa:06} find also younger cores in TMC-1 to have o/p ratios lower than 2, while more evolved cores show a statistical o/p ratio. \cite{park:06} discussed different chemical scenarii at the light of the chemical selection rules to explain  the ratios observed in \cite{morisawa:06} without definitive conclusions.
\cite{ferrerasensio:26} found all of the cores in their sample towards the Perseus molecular cloud to have statistical  c-C$_{3}$H$_{2}$ o/p ratios except for one core which had an o/p ratio even larger than 3. No correlation between the c-C$_{3}$H$_{2}$ o/p ratio and evolutionary state was found. Finally, the spectral survey conducted by \cite{hsu:26} in the Q/W bands with the Yebes 40m telescope found a c-C$_{3}$H$_{2}$ o/p ratio of 2.0$^{+0.7}_{-0.7}$, averaged over 21 sources in the Orion molecular cloud.\

In these cold interstellar environments and particularly in prestellar cores centres \citep{bergin:07, keto:08}, where CO depletion occurs, the H$_{2}$D$^{+}$/H$_{3}^{+}$ ratio is magnified and deuterium is transferred to other molecules resulting into deuterium fractionation, defined as the enhancement of molecular deuterium abundances with respect to the local interstellar medium (ISM) D/H value of 2.0$\pm$0.1 $\times$ 10$^{-5}$ \citep{linsky:03, prodanovic:10, caselli:12, ceccarelli:14}. The deuterated isotopologues  of c-C$_{3}$H$_{2}$, c-C$_{3}$HD and c-C$_{3}$D$_{2}$, are also detected towards different ISM clouds \citep{bell:86, gerin:87, spezzano:13, gratier:16, majumdar:17, chantzos:18, agundez:2019, yoshida:2019, lis:25, ferrerasensio:26}. The doubly-deuterated isotopologue of c-C$_{3}$H$_{2}$, c-C$_{3}$D$_{2}$, also shows separate ortho and para states. In this case, as D has a nuclear spin of 1, the total nuclear spin, $I_{\rm tot}$, can be 0, 1 or 2, corresponding to 1, 3 and 5 values of the statistical weights respectively. Six symmetric nuclear wavefunctions, corresponding to ortho states, arise from $I_{\rm tot}$ = 0 and 2. On the other hand, three anti-symmetric wave functions arising from $I_{\rm tot}$ = 1, correspond to para states. Contrarily to c-C$_{3}$H$_{2}$, the ground level with lowest energy in c-C$_{3}$D$_{2}$ is ortho, and it is then more appropriate to refer to the upper-to-lower ratio as the para-to-ortho ratio (p/o) \citep{flower:04,flower:06,sipila:10}. The states with even $K_a$ + $K_c$ are ortho, and the ones with odd $K_a$ + $K_c$ are para. The lowest para $J_{KaKc} = 1_{01}$ level is only 1.82 K above the ground  $J_{KaKc} = 0_{00}$ ortho level. Consequently, the equilibrium thermodynamic para-to-ortho ratio of c-C$_{3}$D$_{2}$, determined by evaluating the full rotational partition functions for both nuclear spin modifications, is already 0.496 at T = 3 K. Under these conditions, the thermodynamic ratio is essentially equivalent to its nuclear-spin statistical value, making it convenient to refer to this fixed 1/2 value as the 'statistical ratio' throughout this work.\

\cite{spezzano:13} report the detection of one ortho and two para c-C$_{3}$D$_{2}$ transitions towards the L1544 and TMC1 prestellar cores. Nevertheless, they do not report the p/o ratio, as separate ortho and para column densities were not calculated. Similarly, \cite{chantzos:18} reports the detection of both ortho and para c-C$_{3}$D$_{2}$ transitions towards two prestellar sources, L1517B and TMC-2, also without reporting the p/o ratio. Most of the other works in the literature which target starless and prestellar cores commonly only report either ortho or para transitions not allowing to investigate their p/o ratios. Thus, the p/o ratio of c-C$_{3}$D$_{2}$ has not thoroughly been studied to date. Moreover, to our knowledge, no para-to-ortho ratios have been studied for any doubly deuterated molecule with non Local Thermodynamic Equilibrium methods. \ce{D2H+} \citep{vastel:06} and \ce{D2O} \citep{coutens:2014} have only been detected through a single para transition. Other studies on detected doubly deuterated molecules with ortho / para isomers, \ce{ND2H} \citep{roueff:2000}, \ce{D2CO} \citep{ceccarelli:2002} have focused on the deuteration ratios and not been 
considered with this objective. \

The formation pathways of  c-C$_{3}$D$_{2}$ has thus not been studied in detail, so far. \cite{spezzano:13} invoke a cycle of  deuteron transfer reactions of H$_{2}$D$^{+}$ with c-C$_{3}$H$_{2}$, c-C$_{3}$HD and c-C$_{3}$D$_{2}$ followed by dissociative recombination reactions, but deuteration may also proceed through reactions with D$_{2}$H$^{+}$ and even D$_{3}^{+}$ in highly depleted regions \citep{roueff:05}. The electron recombination of the intermediates c-C$_{3}$H$_{2}$D$^{+}$ and c-C$_{3}$D$_{2}$H$^{+}$ can either result in the formation of c-C$_{3}$H$_{2}$, c-C$_{3}$HD and c-C$_{3}$D$_{2}$. Experiments on dissociative recombination of partially deuterated molecular ions (e.g., H$_{2}$D$^{+}$, HDO$^{+}$, HD$_{2}$O$^{+}$; \citealt{datz:95,jensen:99,jensen:00}) have motivated the assumption, commonly adopted in astrochemical models, that channels leading to the ejection of an H atom, and hence the formation of the deuterated neutral molecule may be favoured over those leading to the ejection of D. This behaviour has been used successfully in previous studies of multiply deuterated ammonia chemistry \citep{roueff:05}. It should be noted, however, that the actual branching ratios likely depend on the detailed reaction dynamics and may vary from system to system. \cite{majumdar:17} have considered the impact of the ortho-to-para balance of H$_{2}$, D$_{2}$, H$_{3}^{+}$, H$_{2}$D$^{+}$, D$_{2}$H$^{+}$ and D$_{3}^{+}$
on deuterated carbon molecules in TMC1 and obtain a c-C$_{3}$H$_{2}$/c-C$_{3}$HD ratio about 10 times smaller than the observational result. The scarcity of observational data on c-C$_{3}$D$_{2}$ combined with ortho/para chemistry not yet applied to the modelling of this molecule, has prevented detailed studies of its para-to-ortho ratio.\
 
The present study reports detections of ortho and para c-C$_3$D$_2$ lines in three starless cores located in Taurus. Moreover, we reanalysed previous observations reported in \cite{spezzano:13} and \cite{chantzos:18}, using recently calculated collisional coefficients for c-C$_{3}$D$_{2}$ \footnote{Ben Khalifa et al., to be published}  in the RADEX radiative transfer analysis code \cite{vandertak:07b}. The paper is structured as follows. The observations and the sample are described in Sections \ref{observations} and \ref{sample}, the methods used to analyse the data are reported in Section \ref{analysis}. The results are introduced in Section \ref{results} and their implications are discussed in Section \ref{discussion}. Finally, in Section \ref{conclusions}, we summarise and conclude this work. \

\section{Observations}\label{observations}
The observations presented in this work are part of the Gas phase Elemental abundances in Molecular CloudS (GEMS) IRAM 30m Large Program \citep{fuente:19, navarro:20, rodriguez:21, spezzano:22a, esplugues:22, rodriguez:23}. This program observed 4 different setups (Table B.1 in \citealp{fuente:19}) towards 305 positions located in Taurus, Perseus, and Orion A \citep{rodriguez:21}. The para 3$_{03}$ - 2$_{12}$ and ortho 3$_{13}$ - 2$_{02}$ c-C$_{3}$D$_{2}$ transitions were included in the observed spectral ranges (Table \ref{obsdat}). The observing mode was frequency switching with a frequency throw of 6 MHz well suited for removing standing waves between the secondary mirror and the receivers. The Eight MIxer Receivers (EMIR) and the Fast Fourier Transform Spectrometers (FTS) with a spectral resolution of 49 kHz were used for these observations. The intensity scale is T$_{\rm MB}$, which is related to T$_{\rm A}^{*}$ by T$_{\rm MB}$ = (Feff/Beff)T$_{\rm A}^{*}$ (Table B.1 in \citealp{fuente:19}). 

We also used data from the Yebes 40m telescope (project ID 20A006), in particular the c-C$_{3}$H$_{2}$ ortho fundamental transition (1$_{11}$ - 0$_{00}$) towards B213-C16-1 (Table \ref{obsdat}). Observations were performed in frequency-switching mode using the 7 mm NANOCOSMOS high-electron-mobility transistor (HEMT) receiver, coupled with fast Fourier-transform spectrometers (FFTSs) that provide 8 $\times$ 2.5 GHz bands per linear polarization. This setup covers the 31.3--50.6 GHz frequency range, offering an instantaneous bandwidth of 18 GHz and a spectral resolution of 38 kHz \citep{tercero:21}. A detailed description of these observations, along with a partial analysis of the dataset, is presented in \cite{moral:26}.\

\section{Sample}\label{sample}
\begin{figure}[H]
\centering
\includegraphics[width=8cm]{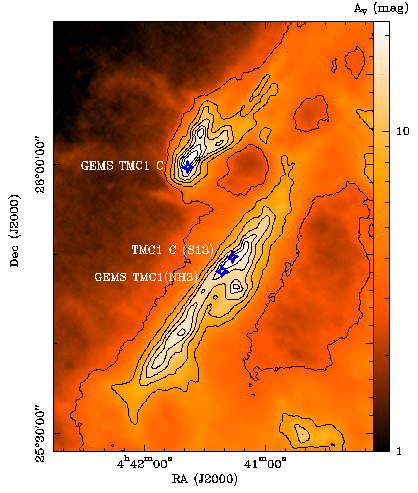}
\caption{The TMC1-C and TMC1-NH3 positions observed in this study as well as the TMC-1C(S13) position studied in \cite{spezzano:13} are marked with blue stars overlaid in an extinction map in colour \citep{fuente:19}. The separation between the TMC1-NH3 and TMC-1C(S13) positions is $\lesssim$ 2$^{\prime}$.} The contours represent 3, 6, 9, 12, 15, 18, and 21 mag.
\label{map}
\end{figure}

In this work we focus on the small subset of the GEMS sources in the Taurus molecular cloud where both ortho and para c-C$_{3}$D$_{2}$ lines are detected with signal-to-noise ratio larger than 3, B213-C16-1, TMC1-C and TMC1-NH3 (Table \ref{source}). Other sources also count with c-C$_{3}$D$_{2}$ detections, but only with either a single ortho or para transitions. Thus, these detections are not suitable for p/o studies, and going to be discussed in the context of c-C$_{3}$H$_{2}$ deuteration on a forthcoming article. These sources are located in the Taurus molecular cloud, a nearby low-mass star-forming region at a distance of 145 pc \citep{yan:19}. Taurus hosts several prominent filamentary structures, including the TMC1 filament, which contains multiple dense cores with well-studied chemical differentiation. In particular, the cyanopolyyne peak (TMC1-CP) and the ammonia peak (TMC1-NH3) mark locations of enhanced carbon-chain and nitrogen chemistry, respectively, and the starless core TMC1-C shows evidence of contraction and accretion motions \citep{agundez:13, feher:16, gratier:16, schnee:07, schnee:10}. We detected the ortho and para c-C$_{3}$D$_{2}$ lines towards positions TMC1-C and TMC1-NH3, which are different from the TMC1 prestellar core position reported by \cite{spezzano:13} (see Figure \ref{map}).\

The filament B213/L1495 in the Taurus molecular cloud has emerged as a key target for recent observational studies (e.g., \citealt{palmeirim:13, hacar:13, marsh:14, tafalla:15, bracco:17, shimajiri:19}). It hosts a rich population of dense cores, which represent a broad range of evolutionary stages, from quiescent, starless condensations to actively star-forming regions associated with young stellar objects (YSOs) \citep{benson:89, onishi:02, tatematsu:04, hacar:13, punanova:18}. A notable characteristic of this region is the clear gradient in stellar surface density along the filament, with a higher concentration in the northern part that progressively declines toward the south. The starless core B213-C16-1 is located in the southern part of the filament, characterized by lower density of YSOs and lower dust temperatures \citep{rodriguez:21, moral:26}.

\begin{table*}[ht]
\begin{center}
\caption{Spectroscopic properties of observed c-C$_{3}$D$_{2}$ transitions.}

\begin{tabular}{ ccccc } 
\hline\hline
 Transition &  & Frequency &  $E_\mathrm{up}$ & $A_{ij}$  \\
 & (o|p) & (MHz) & (K) & ($\times$ 10$^{-5}$ s$^{-1}$)  \\
\hline\hline
 1$_{11}$ - 0$_{00}$ & o & 45358.8400 & 2.18 & 0.39\\
 3$_{03}$ - 2$_{12}$ & p & 94371.3538 & 9.85 & 3.37  \\
 3$_{13}$ - 2$_{02}$ & o & 97761.9779 & 9.89 & 3.88   \\
\hline

\label{obsdat}
\end{tabular}

\tablefoot{The spectral information for the molecular transitions, including the frequency, the upper energy, ($E_{\rm up}$), and the Einstein coefficient, ($A_{ij}$), was taken from the Cologne Database for Molecular Spectroscopy (CDMS)\footnote{\url{https://cdms.astro.uni-koeln.de}}. The c-C$_{3}$D$_{2}$ entry is based on \cite{spezzano:12} high resolution spectroscopy laboratory study. For each transition an o or p indicates whether the transition is ortho or para. The energy origin is that of the ortho 0$_{00}$ level.
}
\end{center}
\end{table*}

\begin{table*}[ht]
\begin{center}
\caption{Physical parameters of the sources targeted in this study. }
\begin{tabular}{ lcccc } 
\hline\hline
 Source & RA & DEC & $n_{\rm H_{2}}$  & $T_{\rm kin}$  \\
 & (J2000) & (J2000) & ($\times$ 10$^{4}$ cm$^{-3}$) & (K) \\
\hline\hline
 B213-C16-1 & 04:21:21.00 & +27:00:09.00 & 3.56$^{+0.26}_{-0.85}$ & 10.9  \\
 TMC1-C   & 04:41:38.80 & +25:59:42.00 & 2.95$^{+2.38}_{-0.55}$ & 11.3  \\
 TMC1-NH3 & 04:41:21.30 & +25:48:07.00 & 2.44$^{+4.91}_{-0.99}$ & 11.7  \\
\hline
\label{source}
\end{tabular}

\tablefoot{The adopted $n_{\rm H_{2}}$ values have been calculated using CCS (Taillard et al., in prep) while the $T_{\rm kin}$ values are derived from \textit{Herschel} maps \citep{rodriguez:21} and assuming that gas and dust are thermalised in these dense cores.}
\end{center}
\end{table*}

\section{Analysis method} \label{analysis}
The fitting of the observed c-C$_{3}$D$_{2}$ transitions is done with the Python ‘pyspeckit’ package \citep{ginsburg:11, ginsburg:22}. The transitions are considered detected if they have a peak intensity $\geq$3$\sigma$, where $\sigma$ corresponds to the noise RMS. This assumption is considered justified as both the main and singly deuterated isotopologues of this molecule (c-C$_{3}$H$_{2}$ and c-C$_{3}$HD) are detected towards these sources. These detections will be presented in a forthcoming article.\

The column densities are calculated using the RADEX code \citep{vandertak:07b}. RADEX is a radiative transfer program designed to simulate non-local thermodynamic equilibrium (non-LTE) excitation conditions. The volume densities of all sources studied in this work, with the exception of L1544, are lower than the critical densities of the c-C$_{3}$D$_{2}$, in the range of 2.6-3.3 $\times$ 10$^{5}$ cm${-3}$ at 5-9 K, so non-thermal excitation is expected. The required input parameters include the volume density ($n_{\rm H_2}$) and kinetic temperature ($T_{\rm kin}$) of the source, as well as the spectroscopic and collisional information of the molecule. As mentioned in the introduction, c-C$_{3}$D$_{2}$ has separate ortho and para levels. The collisional rate coefficients are also separated into the ortho and para species. Thus, ortho and para column densities are computed separately. The collisional rate coefficients of c-C$_{3}$D$_{2}$ have been provided by Ben Khalifa et al. (to be submitted). 

The column densities of the ortho and para c-C$_{3}$D$_{2}$ are determined by comparing observed line intensities to non-LTE radiative transfer models generated with RADEX. For each observed transition, an input file specifying the kinetic temperature ($T_{\rm kin}$), H$_{2}$ density ($n_{\rm H_2}$), background radiation temperature, line width, and a trial molecular column density is constructed. RADEX is then executed to compute the corresponding line flux. The logarithm of the column density is iteratively adjusted to minimize the difference between the observed integrated intensity and the RADEX-predicted flux. Monte Carlo sampling is performed to propagate uncertainties from $T_{\rm kin}$, $n_{\rm H_2}$, as well as the uncertainty on $T_{\rm MB}$, adopted to be the rms of the spectrum, and the uncertainty in FWHM, producing a distribution of column densities from which the median values and confidence intervals are derived. When more than one transition is available for a given spin species, the transitions are fitted simultaneously. This methodology accounts for measurement errors and physical parameter uncertainties.\

Finally, the para-to-ortho ratios are obtained by dividing the para column density by the ortho column density for each source.\

The error propagation for the calculation of the total column densities and para-to-ortho ratios is done with the python package RichValues\footnote{\url{https://github.com/andresmegias/richvalues}}. RichValues is a Python library for working with numeric values that include uncertainties, upper and lower limits, or finite intervals, allowing automatic uncertainty propagation, fitting, and plotting while handling correlations between variables.\

\section{Results} \label{results}

\subsection{GEMS dataset}

\begin{figure*}[h]
\centering
\includegraphics[width=\textwidth]{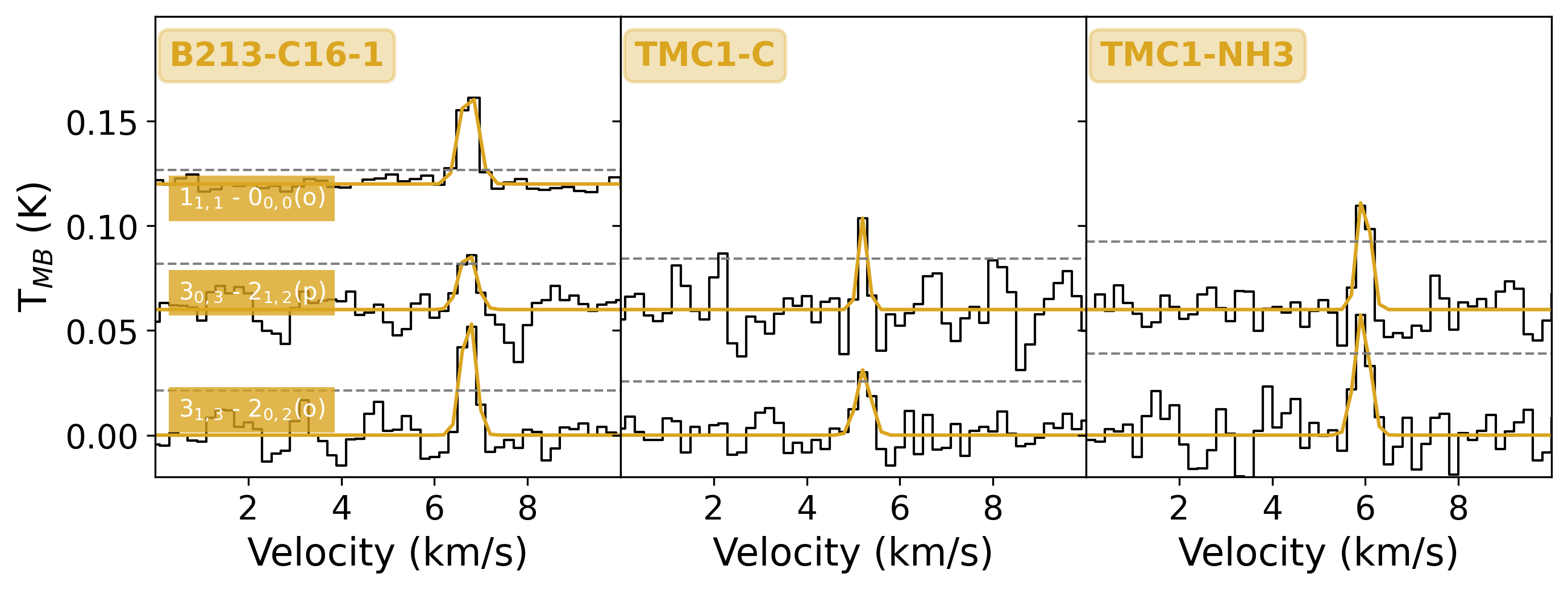}
\caption{Observed 1$_{11}$ - 0$_{00}$, 3$_{03}$ - 2$_{12}$ and 3$_{13}$ - 2$_{02}$ c-C$_{3}$D$_{2}$ lines towards the cores B213-C16-1, TMC1-C and TMC1-NH3 in black. The Gaussian fits are overplotted with a orange line. The horizontal dashed lines show the 3$\sigma$ levels. The different lines observed for each of the cores are shown in the same subplot with a vertical offset of 60 mK.}
\label{occ3d2}
\end{figure*}

The Gaussian fit parameters of the the observed c-C$_{3}$D$_{2}$ lines towards the B213-C16-1, TMC1-C and TMC1-NH3 sources ($T_{\rm MB}$, $V_{LSR}$, FWHM, and noise RMS), as well as the Half-Power Beam Width ($\theta_{beam}$) of the observations are presented in Table \ref{tcc3d2}. Observed lines of c-C$_{3}$D$_{2}$ together with the Gaussian fits to the line profiles are plotted in Figure \ref{occ3d2}.\

\begin{table*}[ht]
\begin{center}
\caption{Observational parameters derived from GEMS detected c-C$_{3}$D$_{2}$ transitions.}

\begin{tabular}{ llcccccc } 
\hline\hline
Source & Transition & $T_{\rm MB}$ & $V_{LSR}$ & FWHM & RMS & S/N & ¥$\theta_{beam}$ \\
 & & (mK) & (km s$^{-1}$) & (km s$^{-1}$) & (mK) & & ($\prime\prime$) \\
\hline\hline
B213-C16-1 & 1$_{11}$ - 0$_{00}$ & 41 & 6.74 (0.01) & 0.44 (0.03) & 2 & 18 & 39 \\
& 3$_{03}$ - 2$_{12}$ & 26 & 6.72 (0.05) & 0.42 (0.13) & 7 & 3 & 26 \\
& 3$_{13}$ - 2$_{02}$ & 52 & 6.73 (0.02) & 0.35 (0.06) & 7 & 7 & 25\\
\hline
TMC1-C & 3$_{03}$ - 2$_{12}$ & 44 & 5.21 (0.07) & 0.23 (0.09) & 11 & 4 & 26\\
& 3$_{13}$ - 2$_{02}$ & 30 & 5.22 (0.05) & 0.36 (0.10) & 8 & 4 & 25 \\
\hline
TMC1-NH3 & 3$_{03}$ - 2$_{12}$ & 50 & 5.97 (0.02) & 0.31 (0.08) & 9 & 6 & 26\\
& 3$_{13}$ - 2$_{02}$ & 57 & 5.93 (0.04) & 0.38 (0.08) & 11 & 5 & 25 \\
\hline
\label{tcc3d2}
\end{tabular}
\end{center}
\end{table*}

Since we have a single transition per spin isomer, except for the B212-C16-1 source, we need to adopt estimations of the the gas kinetic temperature and density obtained from other molecules in previous works for the RADEX calculations. The few detected lines for c-C$_{3}$D$_{2}$ does not allow for the $n_{\rm H_2}$ and $T_{\rm kin}$ parameters to be constrained. Within the GEMS project, the densities have been already estimated using CS and its isotopologues \citep{fuente:19, rodriguez:21}, methanol lines \citep{spezzano:22a}, and more recently from CCS millimetre lines (Taillard et al., in prep). The values obtained from these species are fully compatible within the uncertainties, suggesting that the derived densities are robust (see Appendix \ref{densitiesa}). Based on these results, for the c-C$_{3}$D$_{2}$ calculations, we have adopted the most recent values derived from CCS, due to its closer chemical relationship with c-C$_{3}$H$_{2}$. Following the methodology of previous GEMS papers, we assume that the gas temperature is the same as the dust temperature obtained from \textit{Herschel} maps. The adopted values are shown in Table \ref{source}. \

The ortho and para column densities have been calculated with the 3$_{13}$ - 2$_{02}$ and 3$_{03}$ - 2$_{12}$ transitions, respectively, except for source B213-C16-1 where the ortho column density has been calculated by fitting simultaneously the 1$_{11}$ - 0$_{00}$ and 3$_{13}$ - 2$_{02}$ transitions. We have assumed a beam filling factor of 1 in all the calculations. This assumption has a negligible effect in the cases of TMC1-C and TMC1-NH3 since the frequencies of the two observed transitions are similar. We tried the two options, extended and compact source, to fit the B213-C16-1 observations and obtained the most consistent results with a beam filling factor of 1 (see Appendix~\ref{beama}). \

The column densities derived from the ortho and para lines, as well as the calculated para-to-ortho ratios are summarised in Table \ref{cc3d2otp}. The derived ortho and para column densities are similar within uncertainties for all sources studied in this work. The column densities are of the order of a few 10$^{11}$ cm$^{-2}$. The para-to-ortho ratio is closer to statistical (0.5) for TMC1-NH3, but it is higher for B213-C16-1 and TMC1-C. Nevertheless, the derived p/o ratio for TMC1-C and TMC1-NH3 are subjected to large uncertainties, resulting from the propagation of the $n_{\rm H_2}$ uncertainties.\

The excitation temperatures obtained from the RADEX calculations for the B213-C16-1, TMC1-C and TMC1-NH3, for the ortho and para transitions are: 4.1, 3.8; 4.2, 3.9 and 4.3, 3.0 K, respectively. These excitation temperatures are lower than the kinetic temperatures of the sources, further supporting the sub-thermal excitation scenario and the choice of using RADEX to compute more accurate column densities with respect to other methods which assume LTE. The optical depths retrieved are always lower than 0.05.

\subsection{Para-to-ortho ratios from literature observations}

The coordinates for the sources presented in this section are found in Table \ref{coor} within Appendix \ref{coora}.\

\cite{spezzano:13} report the detection of one c-C$_{3}$D$_{2}$ ortho transition, 3$_{1,3}$ - 2$_{0,2}$, and two para transitions, 3$_{0,3}$ - 2$_{1,2}$ and 2$_{2,1}$ - 1$_{1,0}$ towards the TMC1-C and L1544 cores. The position in TMC-1C observed in \cite{spezzano:13} lays close ($\lesssim$ 2$^{\prime}$) to the TMC1-NH3 position targeted in this study, without overlapping (see Tables \ref{source} and \ref{coor}).
We rename the TMC-1C position from \cite{spezzano:13} as TMC-1C(S13), to differenciate it from the TMC1-C GEMS position, which has a similar label but corresponds to a different position. \cite{spezzano:13} computed the total column densities by assuming LTE conditions with an excitation temperature of 5 K for L1544 and of 7 K for TMC-1C(S13), as no c-C$_{3}$D$_{2}$ collisional rate coefficients 
were available then.\

With the available c-C$_{3}$D$_{2}$ ortho and para collisional rate coefficients, their column densities, as well as their p/o ratios towards L1544 and TMC-1C(S13), may now be reexamined. For this purpose, the observed transitions towards L1544 listed in Table 1 in \cite{spezzano:13} alongside the parameters adopted by \cite{giers:22}, who used an $n_{\rm H_2}$ of 1$\times$10$^{6}$ cm$^{-3}$ and a $T_{\rm kin}$ of 5 K to model the c-C$_{3}$D$_{2}$ 3$_{1,3}$ - 2$_{0,2}$ transition with RADEX. For further discussion on the $n_{\rm H_2}$ and $T_{\rm kin}$ assumptions refer to Appendix \ref{assumpa}. The assumed $n_{\rm H_2}$ and $T_{\rm kin}$ values are also reported in Table \ref{coor}. A 10\% uncertainty for the assumed $n_{\rm H_2}$ and $T_{\rm kin}$ values is taken into account. The ortho 3$_{1,3}$ - 2$_{0,2}$ line is fitted on its own and the para 3$_{0,3}$ - 2$_{1,2}$ and 2$_{2,1}$ - 1$_{1,0}$ lines are fitted together. Similarly to the B213-C16-1 case, the 2$_{2,1}$ - 1$_{1,0}$ transition has a different beam size compared to the 3$_{1,3}$ - 2$_{0,2}$ and  3$_{0,3}$ - 2$_{1,2}$ ones. 
The effect on the calculated p/o ratio by performing beam normalization is explored in Appendix~\ref{beama}, finding that assuming a beam filling factor of 1 gives the best result. The N$_{ortho}$ obtained is 4.6$\pm$0.9 $\times$ 10$^{11}$ cm$^{-2}$ with a $T_{\rm ex}$ of 5~K and an optical depth ($\tau$) of 0.08 and the N$_{para}$ obtained is  2.0$\pm$0.3 $\times$ 10$^{11}$ cm$^{-2}$ with $T_{\rm ex}$ of 5~K and $\tau$ of 0.03 for both transitions. The resulting para-to-ortho ratio is 0.4$^{+0.1}_{-0.1}$. Thus, the c-C$_{3}$D$_{2}$ para-to-ortho ratio measured towards L1544 is compatible with the statistical value within uncertainties for the assumed $n_{\rm H_2}$ and $T_{\rm kin}$.\ 

In the lack of a canonical $n_{\rm H_2}$ and $T_{\rm kin}$ for the TMC-1C(S13) position studied in \cite{spezzano:13} we adopt values in between the ones reported for TMC1-C and TMC1-NH3 of this work (see Table \ref{source}), 2.7$\times$10$^{4}$ cm$^{-3}$ and 11.5 K also assuming a 10\% uncertainty for these values (Table \ref{coor}). The line observed parameters are also taken from Table 1 in \cite{spezzano:13}. As done for L1544, the beam normalisation is tested in Appendix \ref{beama} finding that a beam filling factor of 1 gives the best results. The N$_{ortho}$ obtained is 1.9$\pm$0.2 $\times$ 10$^{11}$ cm$^{-2}$ with a $T_{\rm ex}$ of 4~K and a $\tau$ of 0.07 and the N$_{para}$ obtained is 9.1$\pm$1.3 $\times$ 10$^{10}$ cm$^{-2}$ with $T_{\rm ex}$ of 4~K and $\tau$ of 0.04 for both transitions. The resulting para-to-ortho ratio is 0.5$^{+0.1}_{-0.1}$. Thus, the c-C$_{3}$D$_{2}$ para-to-ortho ratio measured towards TMC-1C(S13) is also statistical within uncertainties for the assumed $n_{\rm H_2}$ and $T_{\rm kin}$.\

The p/o for L1517B is calculated using the observed 3$_{0,3}$ - 2$_{1,2}$ and 3$_{1,3}$ - 2$_{0,2}$ line parameters of Table 3 in \cite{chantzos:18}. The $n_{\rm H_2}$ and $T_{\rm kin}$ values of 2 $\times$ 10$^{5}$ cm$^{-3}$ and 9.5 K, respectively, assuming a 10\% uncertainty were adopted for the calculations \citep{tafalla:02, ford:11}. This resulted in N$_{ortho}$ = 6.8 $\pm$1.8 $\times$ 10$^{10}$ cm$^{-2}$ with a $T_{\rm ex}$ of 7~K and a $\tau$ of 0.01, and N$_{para}$ = 9.5 $\pm$2.5 $\times$ 10$^{10}$ cm$^{-2}$ with a $T_{\rm ex}$ of 5~K and a $\tau$ of 0.02. The resulting p/o ratio is non statistical with a value 1.4$^{+0.7}_{-0.5}$.\

\cite{chantzos:18} reports the detection of three c-C$_{3}$D$_{2}$ transitions, 3$_{0,3}$ - 2$_{1,2}$ (p), 3$_{1,3}$ - 2$_{0,2}$ (o) and 2$_{2,1}$ - 1$_{1,0}$ (p), towards TMC-2 (Table 2 within \cite{chantzos:18}). As for L1544 and TMC-1C, the ortho transition is fitted on its own, while the two para transitions are fitted together. Again, the beam normalization effect test in Appendix \ref{beama}, shows that a beam filling factor of 1 produces the best results. The $n_{\rm H_2}$ and $T_{\rm kin}$ adopted for these calculations are 2 $\times$ 10$^{4}$ cm$^{-3}$ and 10.7 K, assuming a 10\% uncertainty \citep{ford:11}. The N$_{ortho}$ is 4.2 $\pm$0.5 $\times$ 10$^{11}$ cm$^{-2}$ with $T_{\rm ex}$ of 4 K and $\tau$ of 0.14. The N$_{para}$ is 2.7 $\pm$0.6 $\times$ 10$^{11}$ cm$^{-2}$ with $T_{\rm ex}$ of 3 K for both transitions and $\tau$ of 0.06 and 0.09 for 3$_{0,3}$ - 2$_{1,2}$ and 2$_{2,1}$ - 1$_{1,0}$, respectively. The resulting p/o is statistical within uncertainties with a value 0.6$^{+0.2}_{-0.1}$.\ 

Our calculations show that the excitation temperature of the ortho- and para- species can be different, and the p/o ratio derived assuming LTE is lower than the values derived from the non-LTE analysis. 


With the more accurate calculations presented in this paper, we find that only the L1544, TMC-1C(S13) and TMC-2 observations are consistent with a statistical p/o ratio while L1517B shows a possible deviation. However, the uncertainties remain dominated by the assumptions on the physical conditions and the limited number of observed transitions.\

\begin{table*}[ht]
\begin{center}
\caption{Column densities and para-to-ortho ratios of c-C$_{3}$D$_{2}$.}

\begin{tabular}{ lcccc } 
\hline\hline
Source & N$_{ortho}$ & N$_{para}$ & N$_{total}$ & p/o  \\
 & (cm$^{-2}$) & (cm$^{-2}$)& (cm$^{-2}$) &   \\
\hline\hline
B213-C16-1 & 1.1$\pm$0.1 $\times$ 10$^{11}$ &  1.6$\pm$0.7 $\times$ 10$^{11}$ & 2.7$\pm$0.7 $\times$ 10$^{11}$  & 1.5$^{+0.7}_{-0.6}$ \\
TMC1-C & 1.1$\pm$0.5 $\times$ 10$^{11}$ &  1.5$\pm$0.7 $\times$ 10$^{11}$ & 2.6$\pm$0.9 $\times$ 10$^{11}$ & 1.3$^{+1.3}_{-0.7}$ \\
TMC1-NH3 & 2.6$\pm$1.6 $\times$ 10$^{11}$ &  2.7$\pm$1.5 $\times$ 10$^{11}$ & 5.3$\pm$2.2 $\times$ 10$^{11}$ & \textcolor{olive}{1.0$^{+1.4}_{-0.7}$} \\
\hline
L1544$^{a}$ & 4.6$\pm$0.9 $\times$ 10$^{11}$ &  2.0$\pm$0.3 $\times$ 10$^{11}$ & 6.6$\pm$0.9 $\times$ 10$^{11}$  & \textcolor{olive}{0.4$^{+0.1}_{-0.1}$} \\
TMC-1C(S13)$^{a}$ & 1.9$\pm$0.2 $\times$ 10$^{11}$ &  9.1$\pm$1.3 $\times$ 10$^{10}$ & 2.8$\pm$0.2 $\times$ 10$^{11}$  & \textcolor{olive}{0.5$^{+0.1}_{-0.1}$} \\
L1517B$^{b}$ & 6.8$\pm$1.8 $\times$ 10$^{10}$ &  9.5$\pm$2.5 $\times$ 10$^{10}$ & 1.6$\pm$0.3 $\times$ 10$^{11}$  & 1.4$^{+0.7}_{-0.5}$ \\
TMC-2$^{b}$ & 4.2$\pm$0.5 $\times$ 10$^{11}$ &  2.7$\pm$0.6 $\times$ 10$^{11}$ & 6.9$\pm$0.8 $\times$ 10$^{11}$  & \textcolor{olive}{0.6$^{+0.2}_{-0.1}$} \\
\hline
\label{cc3d2otp}
\end{tabular}
\tablefoot{The para-to-ortho values written in \textcolor{olive}{green} are statistical (0.5) within uncertainties. $^{a}$ The column densities for these sources are calculated with the observed line parameters in Table 1 in \cite{spezzano:13}. $^{b}$ The line parameters in Table 3 of \cite{chantzos:18} are used to compute the column densities.}
\end{center}
\end{table*}

\section{Discussion} \label{discussion}

\subsection{The para-to-ortho c-C$_{3}$D$_{2}$ ratio in starless cores}

We present detections of ortho-c-C$_3$D$_2$ and para-c-C$_3$D$_2$ in three starless cores observed within the IRAM 30m large program GEMS (PI: A. Fuente; \citealp{fuente:19}) and the Yebes 40m radiotelescope. These observations allowed us to calculate the c-C$_3$D$_2$ p/o ratio thanks to the recently computed ortho-c-C$_3$D$_2$ and para-c-C$_3$D$_2$ collisional rates (Khalifa et al., to be submitted). In addition, we also evaluate the p/o ratio in 4 additional starless cores \citep{spezzano:13, chantzos:18}. Altogether, we have built a sample of 7 starless cores for which a the p/o ratio is provided for the first time.

The total c-C$_{3}$D$_{2}$ column densities obtained towards B213-C16-1, TMC1-C and TMC1-NH3 are larger than those derived towards a sample of starless and prestellar cores in the Perseus molecular cloud \citep{ferrerasensio:26} by a factor of $\sim$ 3. The column densities reported in this work are however of the same order than those found in L1544 and TMC-C1, reported in \cite{spezzano:13}, and the L1517B and TMC-2 in \cite{chantzos:18}. \

Two out of the three sources from the GEMS dataset appear to show p/o ratios higher than the statistical value. \

The total column density computed in this study for L1544 and TMC-1C(S13) with RADEX using the observations presented in \cite{spezzano:13} are 6.6 $\pm$ 0.9 $\times$10$^{11}$ and 2.8 $\pm$ 0.2$\times$10$^{11}$ cm$^{-2}$, respectively. The column density towards L1544 computed with RADEX agrees within uncertainties with the column density average value computed assuming LTE with the different observed transitions in \cite{spezzano:13}, 6.3 $\pm$ 0.2$\times$10$^{11}$ cm$^{-2}$. The column density value computed with RADEX towards TMC-1C is larger than the value found in \cite{spezzano:13}, 1.4 $\pm$ 0.2$\times$10$^{11}$ cm$^{-2}$, by a factor two. This difference can result from the assumption of the $n_{\rm H_2}$ and $T_{\rm kin}$ values used as input in the RADEX calculations. However, we assume that this non-LTE method provides more accurate constraints on the column densities. For L1517B the total column density 1.6 $\pm$ 0.3 $\times$10$^{11}$ cm$^{-2}$ agrees with the one reported in \cite{chantzos:18}, with the same value. On the other hand, the total column density value for TMC-2, 6.9 $\pm$ 0.8$\times$10$^{11}$ cm$^{-2}$, is larger than the one reported \cite{chantzos:18}, 3.1 $\pm$ 0.3$\times$10$^{11}$ cm$^{-2}$, by a factor of $\sim$2. Similarly to the sources reported in \cite{spezzano:13}, the discrepancy could be due to the calculation assumptions.\

In contrast with previous works \citep{spezzano:13, chantzos:18}, our more accurate RADEX calculations show that 3 out of 4 observed literature positions present p/o ratios consistent with the statistical value of 0.5. Moreover, the obtained p/o ratios do not show any clear correlation with density, which does not support a simple evolutionary trend. Although L1544 is considered a prototype of collapsing core \citep{ward:99, crapsi:05, crapsi:07}, the position TMC1-NH3 in TMC1 is characterized for presenting high abundances of NH$_{3}$ and other molecules related with surface chemistry such as SO and CH$_{3}$OH. Since all the observed positions are located in Taurus, we cannot explore any difference related with star formation activity in the region.\

We are aware that the derived p/o ratios are biased by the assumptions in $n_{\rm H_2}$ and $T_{\rm kin}$ and also by a possible optimistic uncertainty estimates (10\%). The adopted assumptions are considered reasonable taking into account previous studies which targeted these sources. Moreover, when it comes to the assumed $T_{\rm kin}$, due to the similarity in E$_{\rm up}$ amongst the transitions, $\sim$7--10 K, assuming a $T_{\rm kin}$ within the 5--10 K range is not expected to significantly affect the derived p/o ratios. For the L1544, TMC-1C(S13) and TMC-2 sources the para-to-ortho ratios are computed with one ortho transition, 3$_{1,3}$ - 2$_{0,2}$, and two para transitions, 3$_{0,3}$ - 2$_{1,2}$ and  2$_{2,1}$ - 1$_{1,0}$. In the case of B213-C16-1, two ortho transitions, 1$_{1,1}$ - 0$_{0,0}$ and 3$_{1,3}$ - 2$_{0,2}$, and one para, 3$_{0,3}$ - 2$_{1,2}$, are used to compute the p/o ratio. This could add some asymmetry into the uncertainties of the p/o ratio, which depends on the number and excitation conditions of the observed transition. However, we do not detect any trend in our results that could suggest that this asymmetry is affecting the overall conclusions. The p/o ratios calculated for L1544, TMC-1C(S13) and TMC-2 are statistical. The TMC-1C(S13) position is close to the TMC1-NH3 position presented in this work (Figure \ref{map}). The resulting TMC1-NH3 p/o ratio is also calculated to be statistical, showing agreement with the TMC-1C(S13) value derived from \cite{spezzano:13}. On the other hand, the computed p/o for L1517B is higher than the statistical value similarly to the cores TMC1-C and B213-C16-1 belonging to the GEMS sample and observed in a different set of lines.\

For further information on the dependency of the calculated p/o ratio on the assumed $n_{\rm H_2}$ and $T_{\rm kin}$ values, as well as the potential effect of the beam on the results, refer to Appendices \ref{assumpa} and \ref{beama}. \

\subsection{c-C$_{3}$D$_{2}$ para-to-ortho ratio}
\cite{park:06} have studied the chemical routes of formation of para  and ortho c-\ce{C3H2} and the attempts to reproduce the observational ortho-to-para ratios \citep{morisawa:06} are  elusive. The recent observational results of \cite{hsu:26} are also puzzling compared to these predictions. A direct one-to-one correspondence between the ortho-to-para ratio of c-\ce{C3H2}, and the para-to-ortho ratio of c-\ce{C3D2} is not expected, as their respective states are governed by hydrogen and deuterium nuclear spin statistics, respectively, which are fundamentally distinct. Although the ortho-to-para ratio of H$_{2}$ may affect the overall deuteration chemistry \citep{flower:06}, its direct impact on the c-C$_{3}$D$_{2}$ p/o ratio is less straightforward. The study of the para-to-ortho ratio of c-\ce{C3D2} is
thus independent of the previous considerations and requires a new approach.\

Formation of c-C$_{3}$D$_{2}$, as outlined in \cite{spezzano:13}, involves  successive reactions of H$_{2}$D$^{+}$ with c-C$_{3}$H$_{2}$, c-C$_{3}$HD, resulting in c-C$_{3}$D$_{2}$H$^{+}$. The branching ratios expected under nuclear-spin selection rules can be derived assuming full scrambling following \cite{oka:04}:
 	\begin{align*}
  		\ce{H2D+}   + \ce{c-C3HD} &\longrightarrow \ce{\text{o-c-}C3D_2H^+} + \ce{H2} \qquad 2/3  \\
  		\ce{H2D+}   + \ce{c-C3HD} &\longrightarrow \ce{\text{p-c-}C3D_2H^+} + \ce{H2} \qquad 1/3.  \\
    \end{align*}  
The c-C$_{3}$D$_{2}$H$^{+}$ ions are assumed to conserve their para-to-ortho character upon dissociative recombination. Under these assumptions, this mechanism leads to the statistical p/o ratio of 1/2. It is worth noting that, even if this work is focused on cyclic C$_{3}$D$_{2}$, the properties derived from the total nuclear-spin rules detailed above, apply the same way for linear C$_{3}$D$_{2}$.\

p-D$_{2}$H$^{+}$ has been detected at an unexpected large abundance and even mapped in the IRAS 16293E prestellar core \citep{vastel:2004,pagani:2024} and may also contribute to c-\ce{C3D2} chemistry. We tentatively extend the previous schema and consider full scrambling followed by HD ejection, applying the spin selection rules used in the UGAN chemical network \citep{hilyblant:2018}. Although detailed reaction dynamics calculations are still lacking, the assumption of full scrambling may be reasonable for this closed-shell system. The resulting branching ratios are:
 \begin{align*}
   \ce{\text{o-}{D_2H^+}}  + \ce{c-C3HD} &\longrightarrow \ce{\text{o-c-}C_3D_2H^+} + \ce{HD} \qquad 21/27  \\
	\ce{\text{o-}{D_2H^+}}   + \ce{c-C3HD} &\longrightarrow \ce{\text{p-c-}{C_3D_2H^+}} + \ce{HD} \qquad 6/27  \\
   \ce{\text{p-}{D_2H^+}}  + \ce{c-C3HD} &\longrightarrow \ce{\text{o-c-}C_3D_2H^+} + \ce{HD} \qquad 12/27  \\
	\ce{\text{p-}{D_2H^+}}   + \ce{c-C3HD} &\longrightarrow \ce{\text{p-c-}{C_3D_2H^+}} + \ce{HD} \qquad 15/27.  \\
    \end{align*}  
 
The c-C$_{3}$D$_{2}$H$^{+}$ molecular ions are then assumed to dissociatively recombine into c-C$_{3}$D$_{2}$ while preserving their para-to-ortho character. Under these assumptions, if these reactions dominated the formation of c-\ce{C3D2}, and the p/o ratio of \ce{D2H+} was very high, the p/o ratio of c-\ce{C3D2} could approach (5/4).

The possible additional channel introduced by \cite{dias:23} involving direct D-atom incorporation in carbon radicals is not expected to strongly affect the p/o ratio.

\begin{figure}[h]
\centering
\includegraphics[width=9cm]{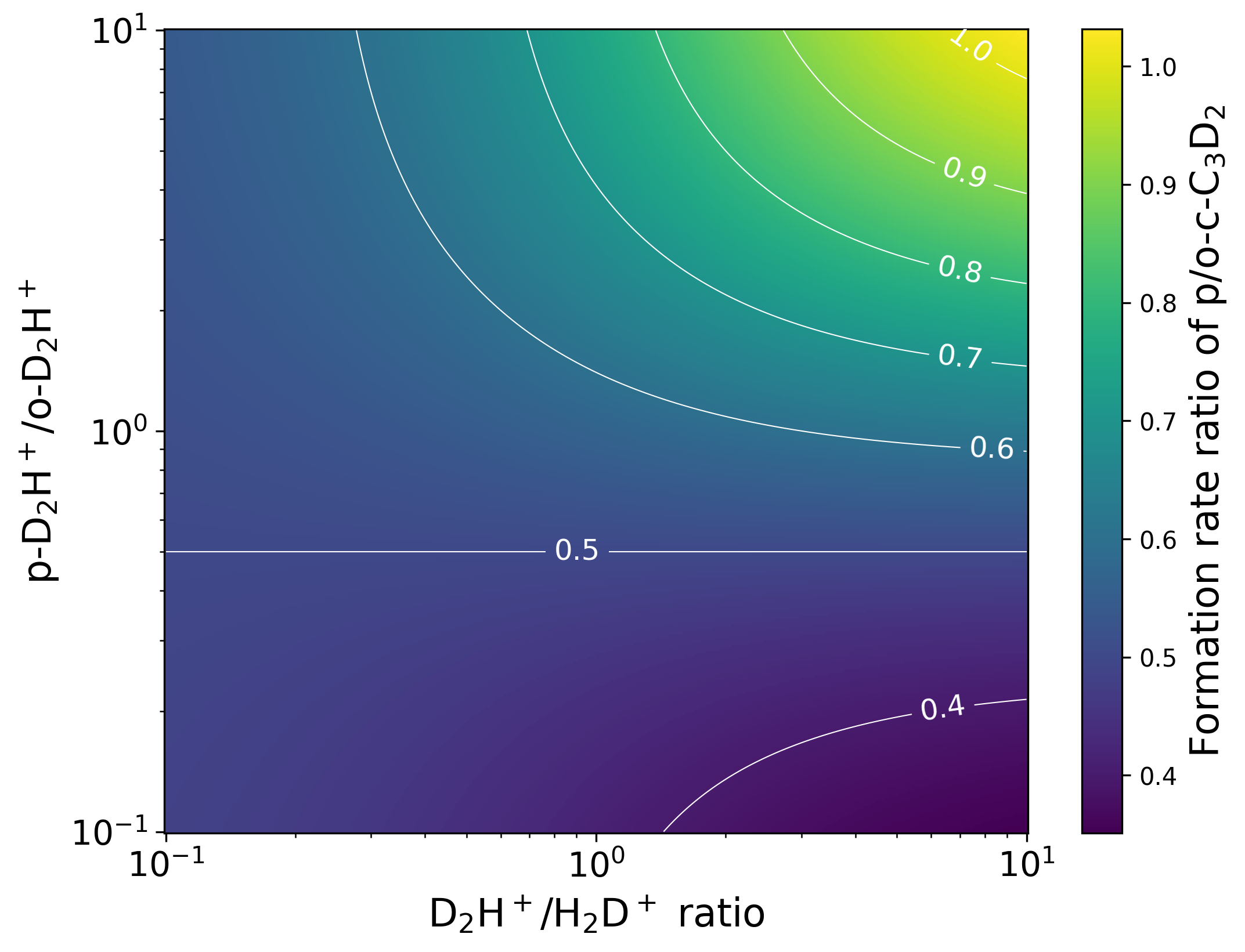}
\caption{Formation rate para-to-ortho ratio of c-C$_{3}$D$_{2}$ in colour scale as a function of D$_{2}$H$^{+}$/H$_{2}$D$^{+}$ and p-D$_{2}$H$^{+}$/o-D$_{2}$H$^{+}$ ratios, in the x and y axes, respectively. The white contours indicate the different c-C$_{3}$D$_{2}$ p/o ratio levels.}
\label{ptogrid}
\end{figure}

With the aim of exploring the c-C$_{3}$H$_{2}$ p/o ratio as a function of the precursor D$_{2}$H$^{+}$/H$_{2}$D$^{+}$ and p-D$_{2}$H$^{+}$/o-D$_{2}$H$^{+}$ ratios, the branching ratios of the reactions introduced in this section are tested within a 0.1 to 10 range for both ratios (Figure \ref{ptogrid}). The results show that to produce p/o ratios larger than the statistical value of 0.5, a combination of high D$_{2}$H$^{+}$/H$_{2}$D$^{+}$ and p-D$_{2}$H$^{+}$/o-D$_{2}$H$^{+}$ ratios are required. In the extreme case where the p/o ratio escalates even further to exceed a value of 1, these chemical conditions appear necessary. Notably, under such high D$_{2}$H$^{+}$/H$_{2}$D$^{+}$ conditions, D$_{3}^{+}$ is expected to be more abundant than D$_{2}$H$^{+}$ and H$_{2}$D$^{+}$ (see Figure 2 in \citealt{walmsley:04}), which indicates that D$_{3}^{+}$ could have a role in the high p/o ratios observed. 

Although, chemical modelling in \cite{sipila:10} shows that in dark cloud conditions the p/o of D$_{2}$H$^{+}$ ranges between 0.1 and 0.4, recent observations by \cite{pagani:2024} shows a higher-than predicted p-D$_{2}$H$^{+}$/o-H$_{2}$D$^{+}$ ratio, 8.3 at 8K, towards the prestellar core IRAS 16293E. The fact that p-D$_{2}$H$^{+}$/o-H$_{2}$D$^{+}$ is higher than predicted can mean that either p-D$_{2}$H$^{+}$ is underestimated, or o-H$_{2}$D$^{+}$ is overestimated in chemical models. In the case p-D$_{2}$H$^{+}$ is underestimated, the D$_{2}$H$^{+}$ p/o ratio can also be higher than expected. 

\section{Conclusions}\label{conclusions}
We have extended the detection of c-C$_{3}$D$_{2}$ \citep{spezzano:13, chantzos:18, ferrerasensio:26} to three additional clouds inluding B213-C16-1, TMC1-C and TMC-NH3 within the Taurus molecular cloud with both para and ortho spin isomers. The separate ortho and para column densities of c-C$_{3}$D$_{2}$ are computed with the non-LTE software RADEX thanks to recent collisional excitation rate coefficients provided by M. Ben Khalifa and J. Loreau leading to non-statistical para-to-ortho ratios for B213-C16-1 and TMC1-C, and a closer to statistical p/o ratio for TMC1-NH3. Moreover, ortho- and para-c-C$_{3}$D$_{2}$ measurements in previous works towards L1544, TMC-NH3-peak, L1517B and TMC-2 are revisited to provide their p/o ratios computed with RADEX. The p/o ratios are close to the statistical value for L1544, TMC-NH3-peak and TMC-2, whereas for L1517B the obtained ratio is significantly greater than 0.5. These results provide a crucial benchmark for future chemical modelling for the formation of c-C$_{3}$D$_{2}$. Specifically, the non-statistical ratios observed suggest that the non-statistical c-C$_{3}$D$_{2}$ p/o ratios may involve reactions with D$_{2}$H$^{+}$ and possibly D$_{3}^{+}$, pathways that could account for the observed para-to-ortho deviations. Yet, the observed p/o values derived in this study present large uncertainties. Further observational multi-transition studies of c-C$_3$D$_2$ in starless cores would help improve the accuracy of the p/o observed values and shed some light on different factors (e.g., environment and time evolution) that could explain the observational results. \

To our knowledge, the present study thus reports for the first time para-to-ortho ratios on a doubly deuterated molecule  derived through the detailed RADEX radiative transfer code.\

\begin{acknowledgements}
J. Ferrer Asensio thanks RIKEN Special Postdoctoral Researcher Program (Fellowships) for financial support. We thank M. Ben Khalifa for sending the updated collision rate coefficients of c-C$_{3}$H$_{2}$, c-C$_{3}$HD and c-C$_{3}$D$_{2}$ in advance of publication. E. Roueff thanks the support of the Thematic Action "Physique et Chimie du Milieu Interstellaire" of INSU Programme national "Astro", with contributions from CNRS Physique \& CNRS Chimie, CEA, and CNES. This project has been carried out with observations from the 30-m radio telescope of the Institut de radioastronomie millimétrique (IRAM) within project 006-17. IRAM is supported by INSU/CNRS (France),
MPG (Germany), and IGN (Spain). Data from the 40-m radio telescope of the National Geographic Institute of Spain (IGN) at Yebes Observatory are also used. Yebes Observatory thanks the ERC for funding support under grant ERC-2013-Syg-610256-NANOCOSMOS. This study is supported by a Grant-in-Aid from the Ministry of Education, Culture, Sports, Science, and Technology of Japan (JP20H05844, JP20H05845 and JP25H00676) and a pioneering project in RIKEN (Basic Science in Space Bridged by Cutting-Edge Space Utilization Technology). A.F. is grateful to the European Research Council (ERC) for funding under the Advanced Grant project SUL4LIFE, grant agreement No101096293. A.F. and B.T. also thanks project PID2022-137980NB-I00 funded by the Spanish Ministry of Science and Innovation/State Agency of Research MCIN/AEI/10.13039/501100011033 and by “ERDF A way of making Europe". B.T. also thanks the Spanish MICIU for funding support from grant PID2023-147545NB-I00.

\end{acknowledgements}

\bibliographystyle{aa}
\bibliography{2026b}

\appendix

\section{Previous estimates of $n_{\rm H_2}$ in our sample}\label{densitiesa}

The H$_{2}$ volume densities for sources B213-C16-1, TMC1-C and TMC1-NH3 in the GEMS sample derived from CS and its isotopologues \citep{fuente:19, rodriguez:21}, from CH$_{3}$OH \citep{spezzano:22} and CCS (Taillard et al., in prep.) are presented in Table \ref{densities}. The density values derived from the different molecules are weakly sensitive to the  selected test molecule.
The $n_{\rm H_2}$ used in this work is the one derived from CCS, due to the chemically related carbon-chain nature of CCS and c-C$_{3}$H$_{2}$.

\begin{table*}[h]
\begin{center}
\caption{Estimated densities for the GEMS sample.}

\begin{tabular}{ lcccc } 
\hline\hline
Source & $n_{\rm H_2}^1$  (cm$^{-3}$)   & $n_{\rm H_2}^2$   (cm$^{-3}$)  & $n_{\rm H_2}^3$  (cm$^{-3}$)   \\
\hline\hline
B213-C16-1   &  2.1$\pm$1.8 $\times$ 10$^{4}$ &  5.1$_{- 1.0 }^{+ 0.6 } \times 10^{4}$    &   $ 3.6 _{- 0.8 }^{+ 0.3 }\times 10^{ 4 }$  \\
TMC1-C       &  4.6$\pm$3.4 $\times$ 10$^{4}$ &  3.8$_{- 0.7 }^{+ 1.0 } \times 10^{4}$    &   $ 2.9 _{- 0.5 }^{+ 2.4 }\times 10^{ 4 }$  \\
TMC1-NH3     &  2.0$\pm$1.0 $\times$ 10$^{4}$ &  1.5$_{- 1.0 }^{+ 0.8 } \times 10^{4}$    &   $ 2.4 _{- 1.0 }^{+ 4.9 }\times 10^{ 4 }$  \\ \hline
\label{densities}
\end{tabular}
\tablefoot{$^1$ Densities derived from CS and its isotopologues \citep{fuente:19, rodriguez:21}. $^2$ Densities derived from CH$_3$OH lines \citep{spezzano:22a}. $^3$ Densities derived from CCS (Taillard et al., in prep)}
\end{center}
\end{table*}

\section{Beam normalization effect on the p/o ratios}\label{beama}

The observational beam sizes of the 1$_{1,1}$ - 0$_{0,0}$, 3$_{0,3}$ - 2$_{1,2}$, 3$_{1,3}$ - 2$_{0,2}$ and 2$_{2,1}$ - 1$_{1,0}$ lines, at 45, 94, 97 and 108 GHz are: 39$^{\prime\prime}$, 26$^{\prime\prime}$, 25$^{\prime\prime}$ and 23$^{\prime\prime}$, respectively. Drawing conclusions from relative line intensities can be problematic if the source is smaller than the beam. The 3$_{0,3}$ - 2$_{1,2}$ and 3$_{1,3}$ - 2$_{0,2}$ lines have similar beam sizes and the column densities derived from them are considered comparable. On the other hand, the 1$_{1,1}$ - 0$_{0,0}$ and 2$_{2,1}$ - 1$_{1,0}$ lines are observed with different beams due to their different rest frequencies. To assess the effect of the beam size in the calculated p/o ratio for sources which include the 1$_{1,1}$ - 0$_{0,0}$ or 2$_{2,1}$ - 1$_{1,0}$ lines, we normalize the observations to the beam of the 3$_{0,3}$ - 2$_{1,2}$ transition, 26$^{\prime\prime}$. To calculate the scaling factor the relation ($\nu_{ref}$/$\nu$)$^{2}$, where $\nu_{ref}$ is the frequency of the 3$_{0,3}$ - 2$_{1,2}$ transition and $\nu$ is the frequency of the transition to scale. This factor is then multiplied to a factor to account for the different diameters of the telescope, and ultimately is multiplied to the line intensity and the column density is calculated the same way with RADEX. The recalculated p/o ratio for the B213-C16-1 gives a value of 0.8$^{+0.4}_{-0.3}$, which is lower than the value 1.5$^{+0.7}_{-0.6}$ presented in Table \ref{cc3d2otp}, but equivalent within uncertainties. The quality of the model is assessed by the derived p/o uncertainties, as well as the $\chi^{2}$ values for the modelled ortho and para integrated intensities. As a reminder, a $\chi^{2}$$\sim$1, corresponds to a good fit, $\chi^{2}$$>>$1 is a bad fit, and $\chi^{2}$$<<$1 is considered a "suspiciously good fit" which can be caused by the simplicity of the model such as fitting a single transition alone. Note that, the simultaneous fit of only two transitions may be not enough to reach $\chi^{2}$$\sim$1. Therefore, the $\chi^{2}$ values of the non-beam-normalized and beam-normalized results are compared only to determine which provides the best-fit model. The combined $\chi^{2}$ for the ortho fit with the beam normalization, 1.8, is smaller than the one without the beam normalization, 3.8. Even if the non-beam-corrected and beam-corrected p/o values are equivalent, the fact that the $\chi^{2}$ of the beam-corrected p/o value is lower than the one of the non-beam-corrected value could point at the beam-corrected p/o value being more accurate. Nevertheless, this can not be confirmed with the current uncertainties. \

The L1544 p/o ratio has been recomputed, taking into account the beam normalization, to give a value of 0.4$^{+0.3}_{-0.1}$ which is equivalent to the value found in the main text, 0.4$^{+0.1}_{-0.1}$. Nevertheless, the $\chi^{2}$ of the beam normalized column densities are significantly larger, by a factor of 2. The p/o ratio calculated with the beam normalization for TMC-1C(S13) is 0.4$^{+0.1}_{-0.1}$ similar than the value presented in the main text, 0.5$^{+0.1}_{-0.1}$. The $\chi^{2}$ for the para is larger for the beam-normalized model, 33.6, compared to the main text result, 19.6. The beam-corrected p/o ratio for TMC-2 gives a value of 0.6$^{+0.1}_{-0.1}$, which is comparable to the non-corrected value presented on the main text, 0.6$^{+0.2}_{-0.1}$. In this case, the $\chi^{2}$ is also larger for the beam normalized case. In view of the resulting beam-normalized p/o ratios, normalising the observations to one beam does not result in better p/o ratios, as determined by $\chi^{2}$, with respect to the non-normalized calculations. For this reason, the non-normalized results are presented in the main text.\

\section{Coordinates of literature sources}\label{coora}

Table \ref{coor}, lists the coordinates for the additional sources where the c-C$_{3}$D$_{2}$ p/o ratio is evaluated as well as the $n_{\rm H_2}$ and $T_{\rm kin}$ assumed for the RADEX calculations.\ 

\begin{table*}[h]
\begin{center}
\caption{Coordinates and assumed $n_{\rm H_2}$ and $T_{\rm kin}$ for the literature sources.}
\begin{tabular}{ lcccc } 
\hline\hline
 Source & RA & DEC & $n_{\rm H_2}$ & $T_{\rm kin}$  \\
 & (J2000) & (J2000) & (cm$^{-3}$) & (K)  \\
\hline\hline
 L1544$^{a}$ & 05:04:17.21 & +25:10:42.80 & 1.0$\times$10$^{6}$ & 5.0  \\
 TMC-1C(S13)$^{a}$ & 04:41:16.10 & +25:49:43.80 & 2.7$\times$10$^{4}$ & 11.5  \\
 L1517B$^{b}$ & 04:55:18.80 & +30:38:04.00 & 2.0$\times$10$^{5}$ & 9.5  \\
 TMC-2$^{b}$ & 04:32:48.70 & +24:25:12.00 & 2.0$\times$10$^{4}$ & 10.7 \\
\hline

\label{coor}
\end{tabular}

\tablefoot{$^{a}$ The coordinates for L1544 and TMC-1C(S13), in the original paper referred to as TMC-1C, are from \cite{spezzano:13}. $^{b}$ The coordinates for L1517B and TMC-2 are taken from Table 1 in \cite{chantzos:18}. The $n_{\rm H_2}$ and $T_{\rm kin}$ assumptions are discussed in the main text.}
\end{center}
\end{table*}

\section{Impact of the $n_{\rm H_2}$ and $T_{\rm kin}$ assumptions}\label{assumpa}

\begin{figure}[H]
\centering
\includegraphics[width=8cm]{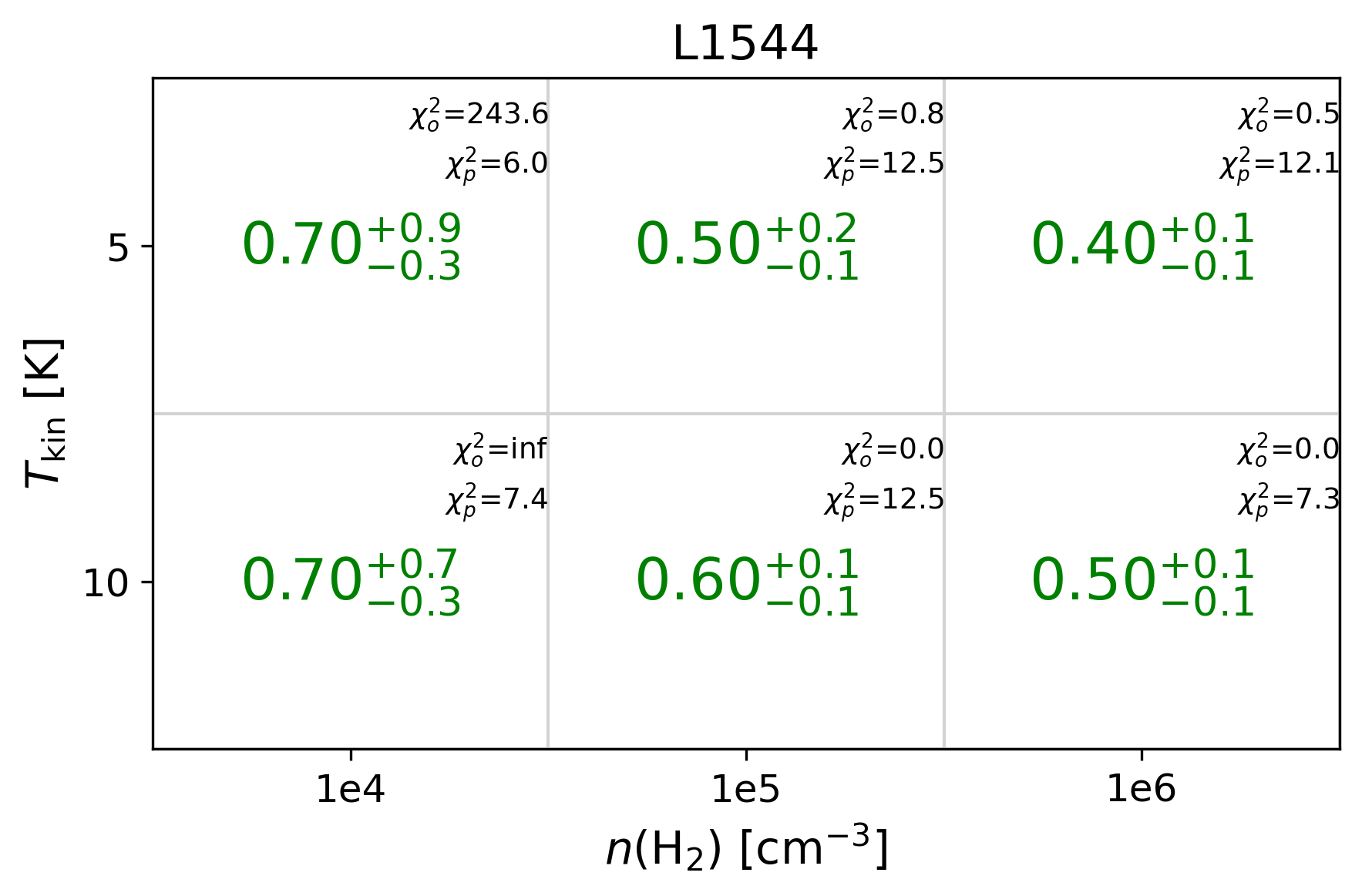}
\caption{c-C$_{3}$D$_{2}$ para-to-ortho ratios for L1544 calculated with the 3$_{1,3}$ - 2$_{0,2}$, 3$_{0,3}$ - 2$_{1,2}$ and 2$_{2,1}$ - 1$_{1,0}$ transitions at different $n_{\rm H_2}$ and $T_{\rm ex}$. As the ortho and para lines are fit separately, distinct $\chi^{2}$ for ortho ($\chi^{2}_o$) and para ($\chi^{2}_p$) are reported. The values in green are in agreement with the statistical value.}
\label{otpgridall}
\end{figure}

With the aim of seeing the dependency of the derived c-C$_{3}$D$_{2}$ p/o ratios to the assumed $n_{\rm H_2}$ and $T_{\rm kin}$ values for sources where these are either not well constrained or not specified in the literature, we use L1544 as a test case. The p/o ratios are calculated for three $n_{\rm H_2}$ values, 1$\times$10$^{4}$, 1$\times$10$^{5}$ and 1$\times$10$^{6}$ cm$^{-3}$, and two $T_{\rm kin}$ values, 5 and 10 K. The uncertainties are assumed to be 10\%. The resulting p/o ratios can be found on Figure \ref{otpgridall}. In our tests we get $\chi^{2}$$<$1 mostly for the fitting of the 3$_{1,3}$ - 2$_{0,2}$ ortho transition. With this in mind, the best fit corresponds to the $n_{\rm H_2}$=1$\times$10$^{6}$ cm$^{-3}$ and $T_{\rm kin}$=10 K. The models assuming 1$\times$10$^{4}$ cm$^{-3}$ give large residuals as well as really large $\chi^{2}$ for the fit of the ortho transition. \

\end{document}